# ANALYSIS OF WIFI AND WIMAX AND WIRELESS NETWORK COEXISTENCE

Shuang Song and Biju Issac

School of Computing, Teesside University, Middlesbrough, UK

*ABSTRACT*

*Wireless networks are very popular nowadays. Wireless Local Area Network (WLAN) that uses the IEEE 802.11 standard and WiMAX (Worldwide Interoperability for Microwave Access) that uses the IEEE 802.16 standard are networks that we want to explore. WiMAX has been developed over 10 years, but it is still unknown to most people. However compared to WLAN, it has many advantages in transmission speed and coverage area. This paper will introduce these two technologies and make comparisons between WiMAX and WiFi. In addition, wireless network coexistence of WLAN and WiMAX will be explored through simulation. Lastly we want to discuss the future of WiMAX in relation to WiFi.*

## 1. INTRODUCTION

With the development of multimedia communication, people need wireless broadband access with higher speed, larger coverage and mobility. The emergence of WiMAX (Worldwide Interoperability for Microwave Access) technology met the people's demand for wireless Internet to some extent. If wireless LAN technology (WLAN) solves the access problem of the "last one hundred meters", then WiMAX technology is the best access solution of the "last mile".

Though WiMAX is an emerging and extremely competitive wireless broadband access technology, the development prospects of its market is still unknown. Hybrid networks as a supplement to cell based or IP packet based services, can fully reflect the characteristics of wide network coverage. It means making a wireless coexistence of Wireless Local Area Network (WLAN or WiFi - a trademarked phrase that means IEEE 802.11x) and WiMAX for devices on different technology segments to communicate with each other.

In this paper the WiFi and WiMAX technologies are introduced initially, and then their own characteristics are compared. Next, the coexistence of WiFi and WiMAX is analyzed. By using the OPNET Modeler software, the wireless coexistence deployment is evaluated with output graphs. Finally, the paper concludes by discussing the future of WiMAX in relation to WiFi.

## 2. OVERVIEW OF WIFI AND WIMAX

### 2.1. WiFi Introduction

WiFi stands for a trademarked phrase which means IEEE 802.11x, and is a short-range wireless transmission technology. WiFi is a technology using wireless means to interconnect personal computers, hand-held devices (such as PDA, smart phone etc.) and other terminals. It is a brand of wireless network communication technology which is held by the WiFi Alliance. The purpose is to improve the interoperability between wireless network products based on the IEEE802.11 standards. Generally, to set up a wireless network, an access point (AP) and wireless network adapters are the basic necessity. This way it can use the wireless medium and coordinate with the structure of the existing wired network to share network resources. As a result the cost of the set up and the complexity are far below the traditional wired network.

Banerji and Chowdhury's study (2013), shows that the coverage area of normal AP is around 20 meters indoors and 100 meters outdoors. It is more suitable for being used in office and home environments. WiFi was the first widely used and deployed high-speed wireless technology, and it has a wide range of applications in networks at home, office, and in growing number of cafes, hotels and airports.

### 2.2. IEEE 802.11 Standards

In recent times IEEE 802.11 wireless local area networks (WLAN) have become ubiquitous across the world in the license-free spectrum of 2.4 and 5GHz bands (Thomas *et al.*,2006).802.11 protocol groups are a wireless local area network standard developed by the International Institute of Electrical and Electronics Engineers (IEEE). The 2.4GHz ISM band is adopted by most of the countries in the world. In some countries and regions, the usage situation of 5GHz ISM band is more complicated. The high carrier frequency has a negative effect, making the popularity of 802.11a limited, although it is the first version of the protocol group.802.11a standard was an amendment of the original 802.11 standard, which was approved in 1999.

802.11 standards have a big family, including about 22 types of standards. In the past ten years, IEEE802.11a/b/g were utilized widely. However now the popular usage is for 802.11n standard that operates in the 2.4GHz and 5GHz bands, with speeds of 400 to 600Mbps (theoretical value). In terms of coverage, 802.11n uses smart antenna technology, through multiple groups of independent antennas to utilize antenna arrays. It can dynamically adjust the beam to ensure that each user receives stable WLAN signals, and can reduce interference from other signals. As its coverage can be extended to few hundred meters through additional devices, the mobility of WLAN has greatly improved.802.11n mainly combines the optimizing of the physical layer and MAC layer to fully enhance the throughput of WLAN technology. Main physical layer technology involves MIMO (Multiple Input Multiple Output), MIMO-OFDM (Orthogonal Frequency Division Multiplexing), 40MHz wide channels, short guard interval and other technologies to make the physical layer throughput up to 600Mbps.

Since the transmission of information is sent through partitioned slots, not only is a single data flow reduced, but the transmission distance can increase with increased antenna range. As a result MIMO technology can increase the existing wireless network spectrum data transmission speed.

OFDM is a high-speed transmission technology in the wireless environment. The main idea of OFDM technology is that the given frequency domain is divided into orthogonal sub-channels. Each sub-channel uses a sub-carrier to modulate and each sub-carrier transmits in parallel (Charles, 2011).

## 2.3. WiMAX Introduction

WiMAX that stands for World Interoperability for Microwave Access, is a standard based on IEEE 802.16 broadband wireless access metropolitan area technology, and it is an air-interface standard for microwave and millimeter-wave band. WiMAX also known as IEEE Wireless MAN (Metropolitan Area Network), can provide an effective interoperability broadband wireless access method under the MAN of a point to multipoint multi-vendor environment. Wireless mesh networks (WMNs) are widely envisioned to be a key technology to improve the capacity and coverage for wireless broadband access services at reasonable costs in rural areas where wired communication infrastructure is too costly to install (Niyato and Hossain, 2007).

The WiMAX Forum is an industry-led, not-for-profit organization which has hundreds of members, comprising most of the WiMAX operators, component vendors and equipment vendors. It was established in June 2001 to promote and certify wireless broadband equipment based on the IEEE802.16 and ETSI HiperMAN (European Telecommunications Standards Institute High Performance Metropolitan Area Networks)standards by awarding equipment manufacturers' products with the 'WiMAX Forum Certified' label (Pareit *et al.*,2012).

As an emerging wireless communication technology, WiMAX provides high-speed connectivity for the Internet, and can be used to connect 802.11x wireless access hotspots to the Internet. Company or personal LAN can also be connected to a wired backbone line. It can serve as a wireless extension cable and DSL technology, enabling wireless broadband access. The signal coverage of WiMAX technology ups to 50km, this technology can operate data communication within the range of 50km at a very fast speed. WiMAX is not only in North America. Europe has rapidly developed in this technology, and this trend has advanced to Asia. WiMAX is another method of providing the "last mile" broadband wireless connectivity solutions for business and home users. Security in WiMAX can also be an issue (Ehtisham, Panaousis and Politis, 2011; Charles, 2011).

## 2.4. IEEEE 802.16 Standards

The IEEE 802 committee set up 802.16 working group in 1999 to specifically develop broadband wireless access standard. IEEE 802.16 is responsible for developing standards for the wireless interface of broadband wireless access and its associated functions. Refer to figure 1. IEEE 802.16 protocol standard consists of two-layer structure, which defines a physical layer and a MAC layer.MAC layer includes 3 parts: service specific convergence sublayer (CS), MAC common part sublayer (CPS) and privacy sublayer. However, encryption protocol sublayer is optional. IEEE 802.16 physical layer defines two duplex modes: Time Division Duplex (TDD) and Frequency Division Duplex (FDD), and these two methods both use burst data transfer format. This transmission mechanism support adaptive burst business data. Transmission parameters (modulation, coding, transmit power, etc.) can be dynamically adjusted, but requires the MAC layer to help the process.

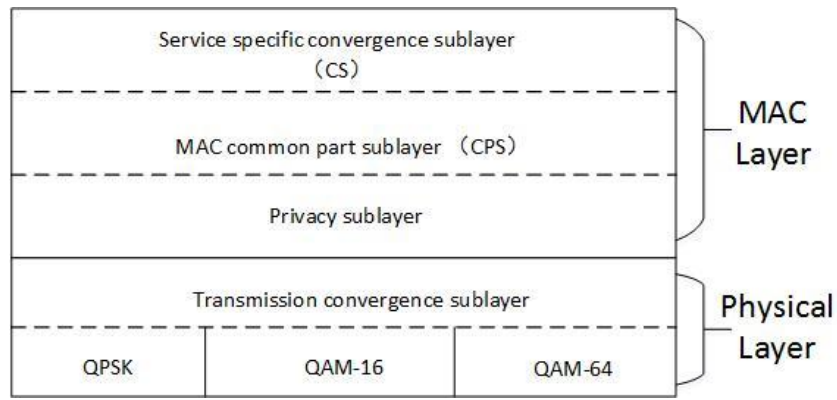

Figure 1. 802.16 protocol stack structure

IEEE 802.16 working group issued IEEE 802.16a in April 2003.This standard supported the work of the band in 2-11 GHz.MAC layer of IEEE 802.16a provided Quality of Service (QoS) assurance mechanism to support voice and video real-time services, added the support for mesh topology network that can adapt to a variety of physical layer environments.

IEEE 802.16e is working in 2 ~ 6 GHz band which supports mobility broadband wireless access air interface standard. IEEE 802.16e is designed to achieve both high-speed data services and enable mobile users with broadband wireless access solutions. In the IEEE 802.16 standard for wireless metropolitan area network (WMAN) in 2004, the IEEE 802.16d standard was published for fixed wireless access (FWA) application. In December 2005, the IEEE ratified the 802.16e amendment, which aimed to support mobile wireless access (MWA) with seamless network coverage. This standard is now receiving considerable industrial attention (Patidar *et al.*, 2012).

Based on IEEE 802.16 series standard, the features of WiMAX are as follows. It achieves around 50 km wireless signal transmission, which cannot be achieved by wireless LAN. The network coverage area is 10 times more than 3G towers. Through the construction of a small number of base stations, the city will be able to achieve full coverage. This makes the wireless network expand the range of applications, providing access speed ups to 70 Mb/s (with 14 MHz carrier).

Wireless standards for different radio links are lot different, causing complexity of varying degrees. For example, network entry and network exit procedures are significantly different in 802.11 and 802.16. 802.11 networks use management frames to do client addition and hand-off, while 802.16 networks use initial network entry procedure (Iyer*et al.*, 2009; EEFOCUS, 2007).

## 2.5. Comparison between WiFi and WiMAX

Some differences between WiFi and WiMAX can be found in the following table. It is quite clear that these two are very different types of technologies with the use of different IEEE standards (Sourangsu, and Rahul, 2013).

Table 1. Comparison of WiFi and WiMAX

| Feature | WiFi | WiMAX |
| --- | --- | --- |
| Standard | 802.11a/b/g/n | 802.16d/e |
| Data rate (MAX) | 300 Mbps | 70Mbps |
| Transmission distance (MAX) | 300m | 50Km |
| Operating Frequency | 2.4 GHz and 5GHz | 2-11 GHz |
| Channel Bandwidth | 20 to 25MHz | Ranging from 1.25 to 20 MHz |
| Encryption | RC4 and Advanced Encryption Standard (AES) | Triple Data Encryption Algorithm (3 DES) and Advanced Encryption Standards (AES) |

## 3. COEXISTENCE OF WIFI AND WIMAX

### 3.1. Deployment of WiFi

WiFi is a wireless network technology that is based on IP addressing. With its high-bandwidth, it has options for mostly small distance communications. As a result, it is mainly used for small-range wireless communications, which is defined as a wireless local area network to meet the requirements of different wireless users.

The simple deployment of WLAN is through an AP or a router to access the Internet. The mobile end devices can receive the WiFi signal from the AP or router. The function of the AP or router is to change the wired network into a wireless network and support more users to connect to the Internet. Refer to figure 2 to see a simple deployment of WiFi network, with the wireless devices (laptop, tablet and smart phone) forming connection to the access point, as though it is a wired network.

The access point should always be part of an existing local area network, through which it can connect to Internet thus providing the node's Internet connectivity.

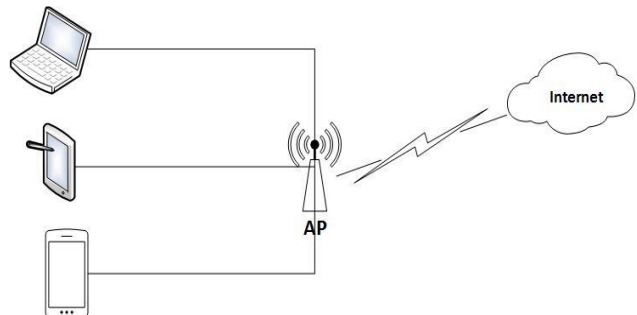

Figure 2. Deployment of WiFi network

## 3.2. Coexistence Deployment

Refer to figure 3 on coexistence network of cellular, WiMAX and WiFi technologies. The cellular technologies (like 3G, as 4G is not that well covered now) lack the bandwidth of WiFi network but has better mobility, while WiFi has better bandwidth with lesser mobility.

These are the shortcomings of WiFi technology when we think of the large wireless coverage:

(1) The characteristics of WiFi determine that APs should have their own channels in an area, or it will cause interference. Among multiple operators, billing and roaming have become a restricting factor in the development.
(2) By the small transmission distance limitation, each WiFi access point becomes a network island. Therefore it is difficult to cover the entire city.
(3) It cannot be used by itself with a high speed moving vehicle such as the metro, railway and bus transit system. So the network would fail to truly support the dream of a mobile city.

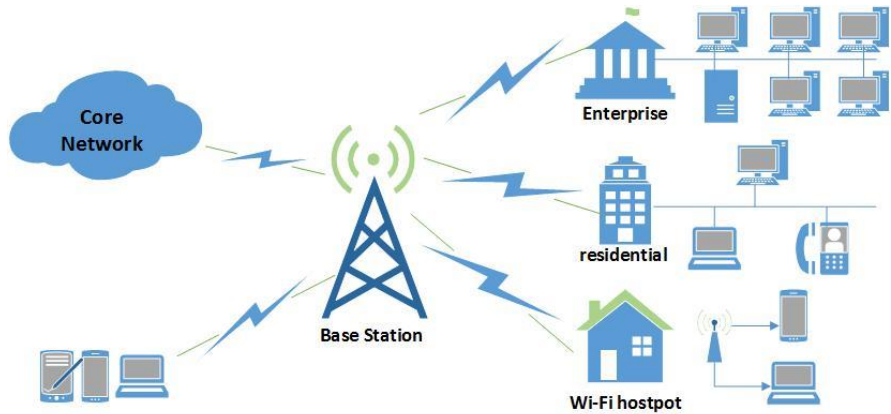

Figure 3. Wireless coexistence (WiFi, WiMAX and 2G/3G)

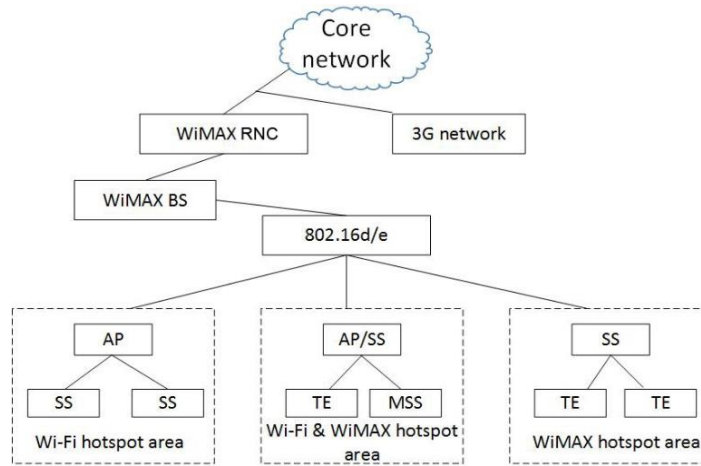

Figure 4. Hybrid network (WiFi and WiMAX)

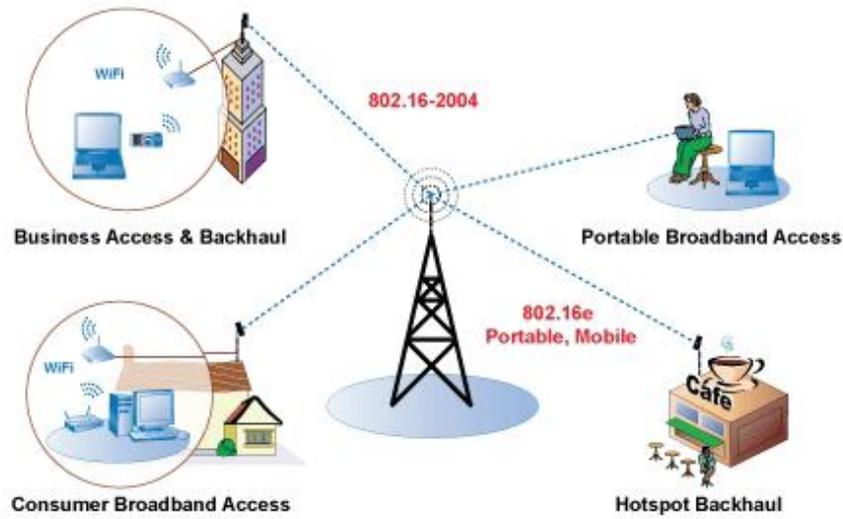

Figure 5. WiMAX network (Shabbir Ahmed, 2014)

The most constructive approach is that WiMAX and WiFi are strongest when working collaboratively. Therefore multi-mode cards (for multi-mode devices) will revolutionize the roaming hotspot user's experience. In addition, the technologies will coexist in a creative way (Marzuki and Baba, 2011).Coexistence of WiMAX and WiFi, can solve the discussed problems of WiFi as in figure 4. The development of WiMAX and WiFi is a complementary trend. In the recent times they have coexisted with each other, and coordinated well with 3G technology. WiMAX, WiFi and 3G joint network make use of a unified management platform to share user's information. Meanwhile, the performance of the existing networks can be greatly improved.

WiMAX (as in figure 5) has been deployed in three phases. The first stage is using an indoor antenna to deploy IEEE802.16d standard WiMAX technology. The target user is the known subscriber in a fixed location. The second stage should deploy substantial indoor antennas. WiMAX technology will broaden the appeal to operators who seek to simplify the user point of installation. The third stage is where the IEEE802.16e standard is launched. WiMAX certification hardware in this standard will be used in portable solution for those users who want to roam in the service area, supporting similar to today's WiFi capabilities, but with more persistent and stable connection.

Gumaidah, B. F., Soliman, H. H., and Obayya, M. (2012) discusses the performance of WiMAX with voice communication. The results show that the higher the base frequency the higher the Signal to Noise Ratio that leads to high throughput, low packet end to end delay. WiMAX systems are expected to deliver broadband access services to residential and enterprise customers in an economical manner (Shabbir Ahmed, 2014).

Leaving aside cellular networks, a practical way of having WiMAX and WiFi joint network is to use WiMAX to link up WiFi hotspots. It can provide E1/T1 and IP dual-channel wireless transmission for WLAN AP and achieve a wider range of high-speed wireless access. In this way, WiFi can increase its access region and provide users with better data services.

## 3.3. Coexistence Application

In July 2010, Taipei City Government started wireless Internet service in specific places, then in October it was launched in all outdoor public places. In total there were approximately 540 wireless access points (AP), and about 500 wireless APs on the buses. It offered free WiFi Internet access. The transmission bandwidth was 512Kbps. Taipei City Government Information Office maintained that there were more than 2000 APs in Taipei (Sina, 2011).

The whole Taipei wireless network used WiMAX and WiFi as two transmission technologies. Provisioning architecture was divided into three parts such as - data center, wireless base stations on high buildings and wireless APs in each place. The base stations were set up on the roof of the buildings, through a 50Mbps - 100Mbps fiber network link to the data center, and the signal range of each base station covered a radius from 500 to 1000 meters, transmitting 4Mbps ~ 8Mbps bandwidth WiMAX signals to each corner of the AP. The AP turned WiMAX to WiFi signal to provide WiFi service.

# 4. OPNET MODELER

## 4.1. Network Simulation Overview

Digital communication network has developed for more than 20 years and in this process the communication network has been progressively computerized. Its structure and function have become more complex, and technology has grown faster. As a result, the research and development of communication networks have become more and more difficult. Thus innovative development of the traditional method is required. It needs the research and development methods which is based entirely on the physical entities by using simulation tools as a support or verification mechanism. In addition to the design, any simulation tool can be used to validate, test products, and thus reduce research costs.

Network simulation is a method to simulate the network behaviour using mathematical modelling and statistical analysis. Simulating the transmission of network traffic we can do an optimized network design to check the network data.

Hughes (2009) referred that network simulators attempt to model real world networks in his work. The idea being that if a system can be modeled, then features of the model can be changed and the results analyzed. As the process of model modification is relatively cheap, then a wide variety of scenarios can be analyzed at low cost (relative to making changes to a real network).

## 4.2. OPNET Modeler Introduction

OPNET (Optimized Network Engineering Tool) Company originated in Massachusetts Institute of Technology, and it was established in 1986. In 1987, OPNET company released its first commercial network performance simulation software which provided an important network performance optimization tool which revolutionized network simulation. Making powerful predictive network performance management through simulation has thus become possible. OPNET has developed other products besides Modeler, and it also includes OPNET Development Kit, WDM Guru etc.

International Journal of Computer Networks & Communications (IJCNC) Vol.6, No.6, November 2014

According to Long (2006), there are various types of products in OPNET for different networking needs. This software uses an object-oriented modeling and graphical editors. It reflects the structure of actual networks and network components. It provides comprehensive support communication systems and distributed systems development environment. Flexible hierarchical modeling method of OPNET Modeler can support all network research related communications, devices and protocols (Song, 2010).

### 4.3. Wireless Network Coexistence Deployment

This experiment was conducted using the OPNET Modeler with existing OPNET simulation models and scenarios. We analyzed them as follows with a simple topology that included a mixture of WLAN and WIMAX components. Figure 6 shows the topology used with use of WiMAX technology as the backbone network of a WLAN hotspot.

In a small area, both WiMAX base station (BS) and WLAN-WiMAX access point (AP) are present. WLAN nodes (clients) on the left are connected to the network servers using WiMAX connections.

Some of the WiMAX simulation parameters are as follows:

| Antenna Gain | 14 to 15 dBi |
|---|---|
| Maximum Transmission Power | 0.5 to 3W |
| Receiver Sensitivity | -200 dBm |
| Maximum number of nodes | 100 |
| PHY Profile/Characteristics | OFDM |

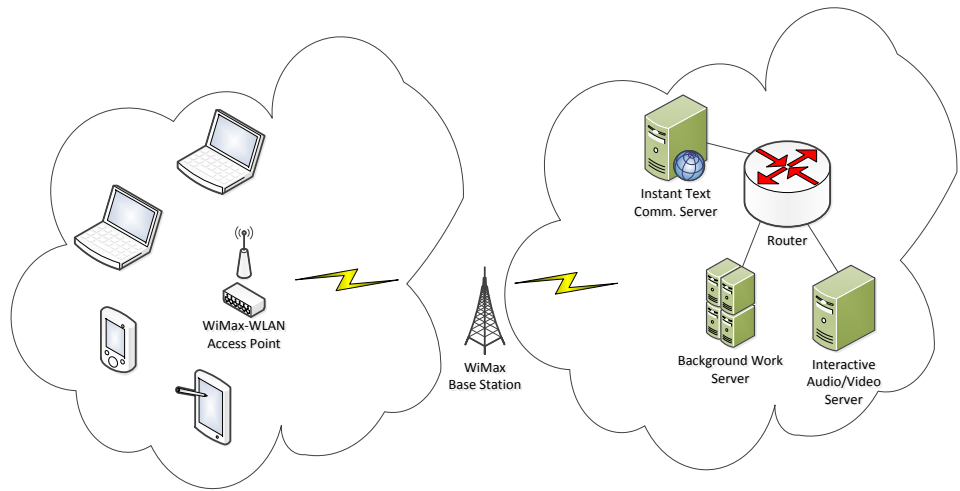

Figure 6. Wireless (WLAN-WiMAX) coexistence topology

Some of the WiFi simulation parameters are as follows:

| | |
|---|---|
| Data Rate | 11 – 54 Mbps |
| Packet reception Power Threshold | -95 dBm |
| CTS-to-self option | enabled |
| AP Beacon Interval | 0.02 sec |
| PHY Profile/Characteristics | Direct Sequence |
| Large packet processing | drop |

### 4.4. Simulation Results and Discussions

The simulation ran for 30 minutes (1800 seconds). The networks used RIP for routing. We measured traffic on the client and server sides. Figures 7, 8 and 9 demonstrates the traffic received by wireless users which contains Background Work Server, Instant Text Communication Server and Interactive Audio/Video Server. Time is represented on x-axis of graphs in seconds.

Some observations we made are as follows: The first graph in figure 7 shows the Work Server is sending a steady traffic (for background processes) at around 80 to 100bytes per second, which is received by the user with some loss. However, the trends of figure 8 and 9 are quite different from figure 7. They reveal a fluctuant and growing traffic; the maximum values of Instant Text Communication Server and Interactive Audio/Video Server are 90 byte/sec and 90,000 byte/sec respectively. The traffic is also dependent on WiFi and WiMAX translations happening. Wireless users can connect the backbone network with the help of WiMAX BS and WLAN hotspot.

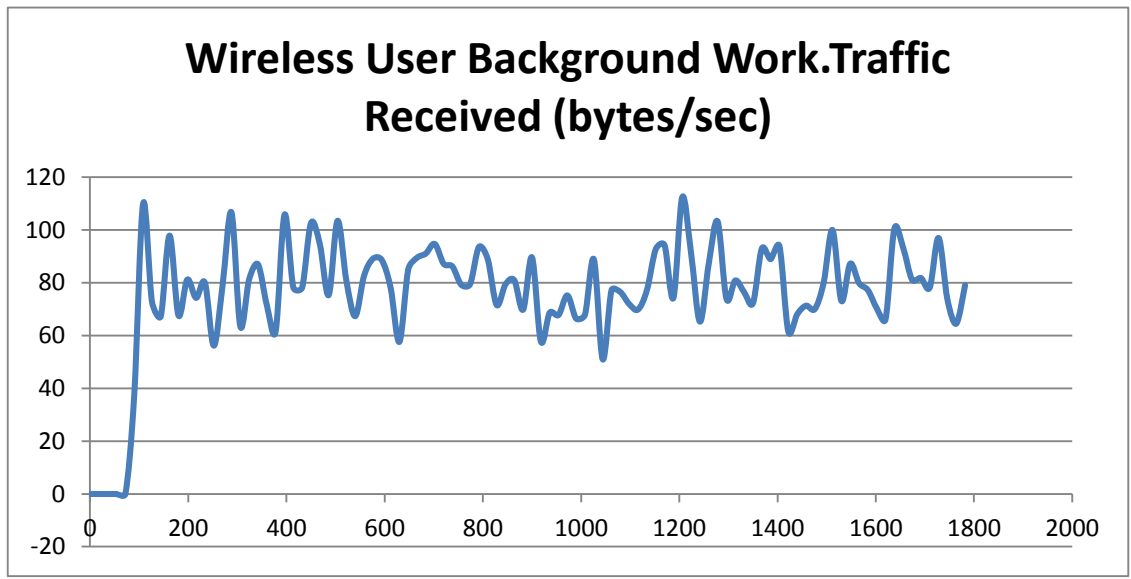

Figure 7. Wireless User Background Work Traffic

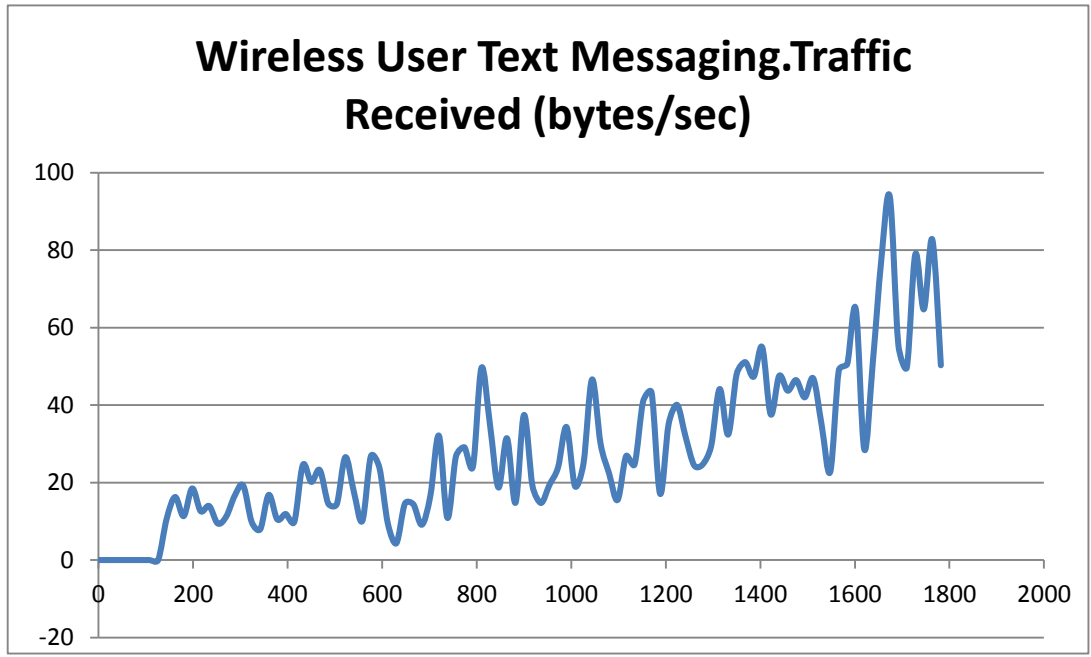

Figure 8. Wireless User Instant Text Traffic

In figure 10, the throughput of WLAN and WiMAX produced the same result measured on either side of WiMAX-WiFi AP. For both of them, the traffic measured is generated from the Interactive Audio/Video Server and the result looks quite reasonable.

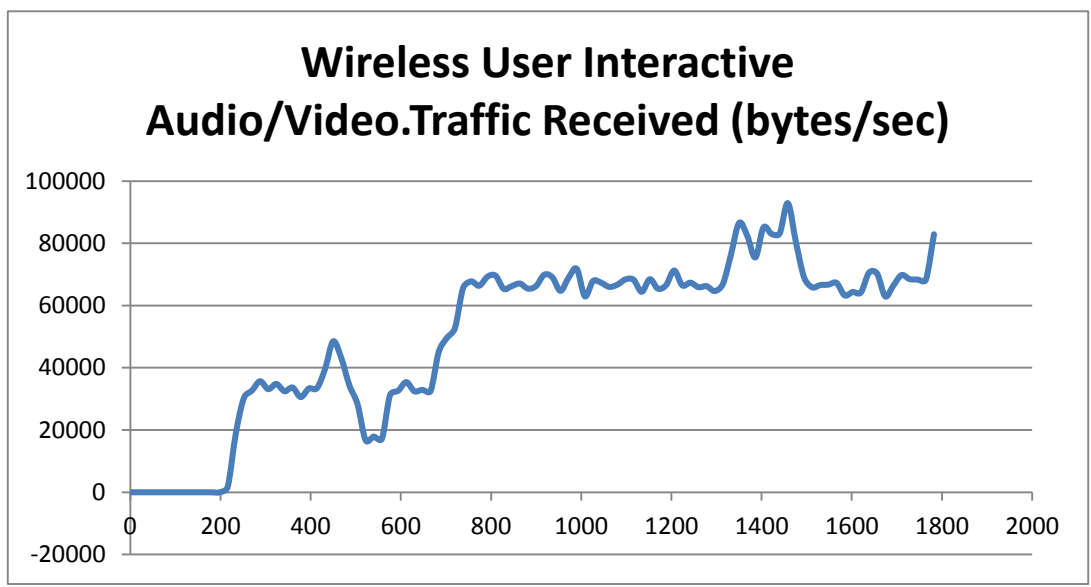

Figure 9. Wireless User Interactive Audio/Video Traffic

International Journal of Computer Networks & Communications (IJCNC) Vol.6, No.6, November 2014

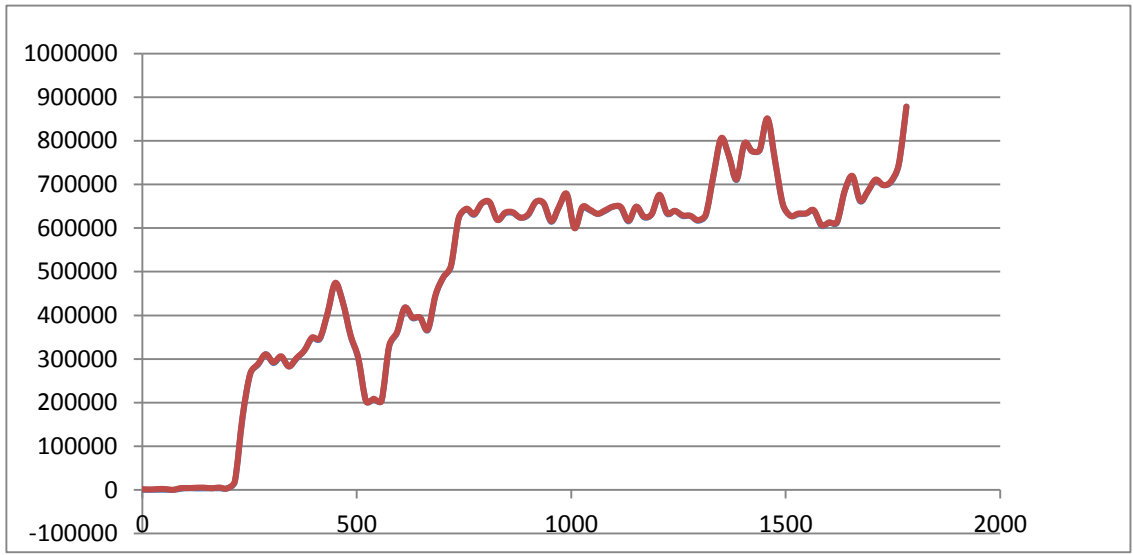

Figure 10. WLAN and WiMAX throughput (bits/sec)

In figure 11 and 12, the traffic delay time of WiMAX on server side is longer than WLAN on client side. The WLAN delay is lesser on the client side with lighter load on them, but it is slightly higher on the WiMAX side servers with higher load on them. The WiMAX tends to perform slightly lower than WLAN when it comes to latency or delay. The average value of WiMAX is 0.05 seconds and the maximum of WLAN is about 0.0017 seconds.

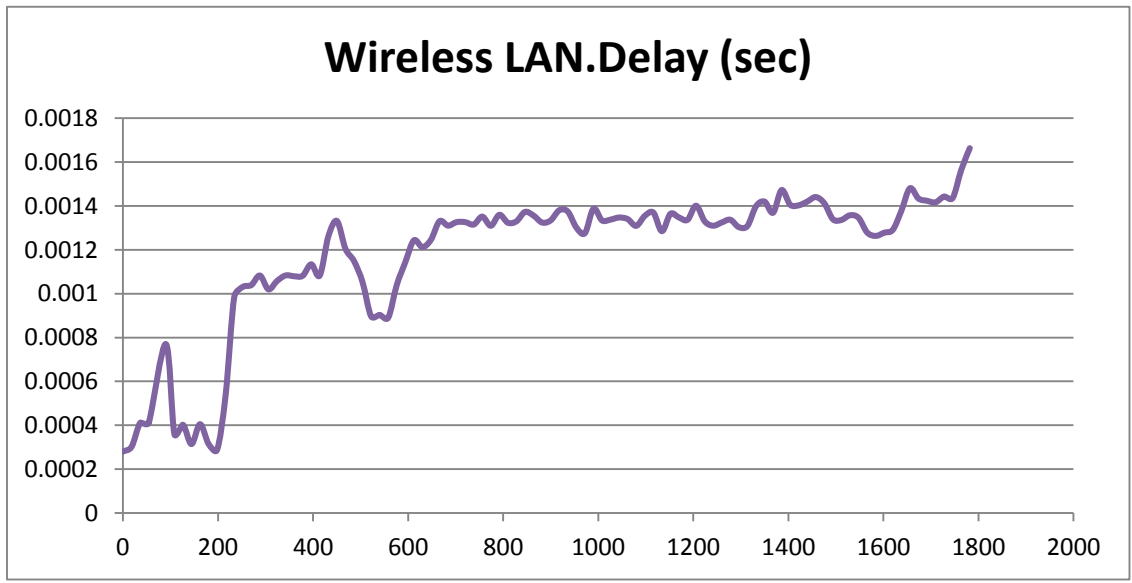

Figure 11. Average WLAN Delay on client side

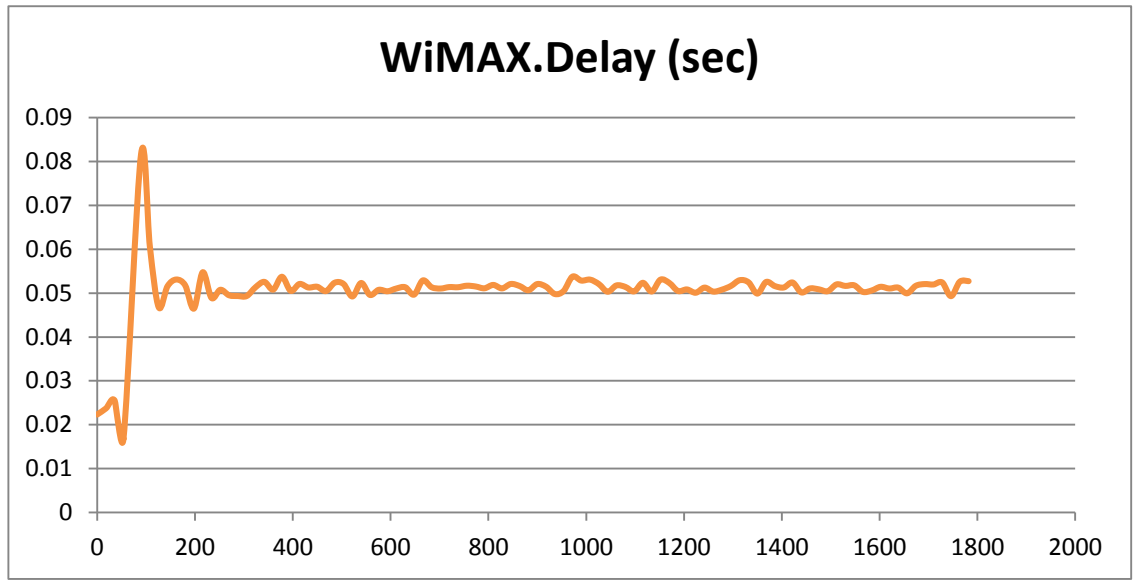

Figure 12. Average WiMAX Delay on server side

## 5. CURRENT SITUATION OF WIRELESS NETWORKS

### 5.1. Trends of Enterprise WiFi Market

As more and more equipment which support 802.11ac emerged, suppliers decrease the cost of 2x2 802.11ac access point (AP). Therefore, the adoption rate of 802.11ac AP will continue to rise. Cloud solutions for WiFi management and services will extend to SMEs (Small and Medium Enterprises) to provide more services to companies' technology. Analysis of WiFi based location in each organization will play an increasingly important role in helping them enhance business intelligence analysis, determine the security policy and enhance customer / user WiFi experience. Analysis of associating the customer's data point (location, application, device type, trends etc.) is particularly important. The solution of BYOD (Bring Your Own Device) and the popularity and integration of device management trend will continue to extend. As certified by Passpoint and ready to enter the smart phone market, Hotspot 2.0 will serve another way to the enterprise providing WiFi access. With social media (Facebook, Google+) as the user login credentials, WiFi will spread the various organization's network reach and could provide guest access.

### 5.2. Operator / Service Provider WiFi Market Trends

In future, large-scale Hotspot 2.0 roaming alliances could be established, automatically and securely connecting to WiFi network which consists of millions of APs and thousands of roaming partners. These partners will include a large number of public places which have indoor WiFi deployment. Companies continue to seek WiFi service providers to deal with many problems, such as network complexity, new services and lack of technologies for many IT organizations. Location Services will bring profit opportunities for the enterprises which provide hosting services. With 802.11r and 802.11k technologies go into mobile devices and AP, seamless WiFi switch will come true. Even if the user enters another coverage area, the smart phone does not need to depend on the relevant AP. With the help of WiFi seamless switch, its experience is comparable to cellular service. Carrier-grade WiFi management system is beginning to catch up with carrier-grade WiFi network infrastructure.

### 5.3. Future Development of WiFi

The Wireless Broadband Alliance (WBA, an industry association aimed at promoting the next generation WiFi experience) in a published report pointed out that by 2015, the number of global public WiFi hotspots will grow by 350%. This figure does not even include "community hotspots" and multiple users share a WiFi access point, while this figure is as high as 4.5 million.

It is expected that China (China Mobile) and Japan will deploy new hotspots (Cheng, 2012), to cover three kinds of places: open outdoor hot zones (like the parks), outdoor hot zone areas (like the tourist attractions) and transportation hub (like the airports). The chairman of the WBA, BT chief executive Chris Bruce said the survey results show that the world is about to usher in the "the Golden Age of public WiFi", and the number of hotspot deployments is expected to soar. The survey also found that more people use the smart phone to link to the WiFi hotspots than laptop users. Laptop users are currently less than half the number of those who access hotspots (48%), while smart phone users accounted for 36%, and fixed PC accounted for 10%.

### 5.4. Development Bottleneck of WiMAX

As the IEEE 802.11n standard has been widely used, the WiFi seems "more like" the WiMAX which has fast transmission speed and wide coverage area, though coverage is better with WiMAX. Back in 2008, the global financial crisis gave indications to the development prospects of WiMAX which was one of the international 3G standards. In the first half of the year, Nortel announced to stop investing in the development of WiMAX product line. After that, Alcatel-Lucent announced that it would reduce investment in WiMAX business. The financial crisis had affected the confidence of investors to put money in the network operators to build WiMAX. Although many countries operators had got WiMAX licenses, they also stopped the construction.

Intel has always been a strong support for WiMAX wireless technology, but in 2010 it announced in an internal meeting that the WiMAX Program Office, which was used for the promotion of technical development has been disbanded. Though in the past years Intel had created a real action-oriented Internet, and WiMAX was one of the key elements, it finally rejected investment through Intel Capital for the 30 global WiMAX technology vendors and service providers.

### 5.5. Current Situation of WiMAX

WiMAX might prove to overcome the 3G LTE (Long Term Evolution) technologies to become the winner of 4G competition, because WiMAX was the first emerging technology. But the truth is that WiMAX is failing and so global operators will focus on the construction of LTE.

In 2014 WiMAX network faced failure again. In March, Taiwan Provincial Government intended to use existing WiMAX business 2600 MHz band to release 190MHz bandwidth for operator's bidding. Currently 2600MHz is used for wireless broadband access services (WBA), which means it is used by WiMAX, but WiMAX operator's spectrum licenses will expire in 2014 and 2015. Then WiMAX operators can choose to change licenses to continue to operate WiMAX services, but the government will increase the license fees. American carriers Sprint also announced that the company planned to close WiMAX network bought from the Clearwire by the end of next year. Sprint announced in a statement the closure of Clearwire's WiMAX network as well, and the company plans to expand its 2.5GHz LTE deployment, which will conduct technology upgrade to about 5000 traditional Clearwire base stations which are expected to be completed by the end of 2015.In most of its 800MHz and 2.5GHz band 4G LTE will be deployed.

4G networks is becoming the mainstream choice of global operators. Some previously deployed WiMAX network operators began to shut down parts of the network or spectrum to free up resources to build 4G LTE networks. LTE is thus getting used in mobile phones and high-speed wireless communication standards. On December 6, 2010 ITU announced LTE Advanced is officially known as 4G.

## 6. CONCLUSION

We could analyze the performance of the co-existence network of WiMAX and WiFi through existing OPNET simulation models and the network was performing as anticipated. The data traffic received and delays were measured. Undeniably the advantages of WiMAX technology in some areas are quite obvious, but the difficulties it faced indeed hindered its further development. If WiMAX wants to continue to develop, it must be redeployed and make a reasonable plan for its market positioning. According to Kamali et al. (2012) WiMAX enables low cost mobile access to the Internet and provides integrated wireless fixed and mobile services using single air interface and network architecture. But the place of WiFi cannot be replaced in the recent years. The better development of WiMAX will be to consider how to coexist with WiFi (Arlene, 2012). Like the deployment picture in this paper, using these two technologies can achieve a better coverage of wireless networks. WiMAX has moved into a bottleneck position that could not be avoided. Because the mobile operators are gradually moving to LTE market, WiMAX had to face enormous challenges. At the same time, this gave WiFi a lot of space to continue the development. Undeniably WiMAX has advantages in some areas, but it still faces the risk of being acquired. WiMAX should pinpoint its location, find the appropriate development of the market, and think of better coexistence with WiFi.

## Authors

Shuang Song has completed MSc Computing from School of Computing, Teesside University, UK. She has studied Network System, Security and Network Service Management and has research interest in Wireless Networks.

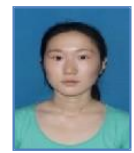

Biju Issac is a senior lecturer in the School of Computing, Teesside University, UK. He has completed Bachelor of Engineering in Electronics and Communication Engineering (ECE), Master of Computer Applications (MCA) with honours and PhD in Networking and Mobile Communications, by research. He is a Charted Engineer (CEng), and Senior Member of IEEE.

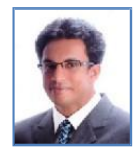